\documentclass[fontsize=11pt, paper=letter]{scrartcl}
\usepackage[utf8]{inputenc}
\usepackage[greek, english]{babel}
\usepackage{authblk}
\usepackage{siunitx}
\usepackage{amsmath}
\usepackage{amssymb}
\usepackage{graphicx}
\usepackage[includeheadfoot, headsep=.1in, margin={.6in,.6in}]{geometry}
\usepackage{hyperref}
\usepackage{array}
\usepackage{textcomp}
\usepackage{float}

\newcommand{\mytilde}{\raisebox{-0.9ex}{\~{ }}}

\begin{document}

\title{Scanning Fluorescence Correlation Spectroscopy (SFCS) with a Scan Path Perpendicular to the Membrane Plane}

\author{Paul Müller$^1$ , Petra Schwille$^2$, Thomas Weidemann$^2$}

\date{August 2013}

\affil{$^1$Biotechnology Center of the TU Dresden, Dresden, Germany}
\affil{$^2$Max Planck Institute of Biochemistry, Martinsried, Germany}

\maketitle

\begin{center}
The final publication \cite{self} is available at \url{http://link.springer.com} or via \url{http://dx.doi.org/10.1007/978-1-62703-649-8_29}.
\end{center}

\abstract{Scanning fluorescence correlation spectroscopy (SFCS) with a scan path perpendicular to the membrane plane was introduced to measure diffusion and interactions of fluorescent components in free standing biomembranes. Using a confocal laser scanning microscope (CLSM) the open detection volume is moved laterally with kHz frequency through the membrane and the photon events are continuously recorded and stored in a file. While the accessory hardware requirements for a conventional CLSM are minimal, data evaluation can pose a bottleneck. The photon events must be assigned to each scan, in which the maximum signal intensities have to be detected, binned, and aligned between the scans, in order to derive the membrane related intensity fluctuations of one spot. Finally, this time-dependent signal must be correlated and evaluated by well known FCS model functions. Here we provide two platform independent, open source software tools (PyScanFCS and PyCorrFit) that allow to perform all of these steps and to establish perpendicular SFCS in its one- or two-focus as well as its single- or dual-colour modality.}


\section{Introduction}
In recent years, fluorescence correlation spectroscopy (FCS) has become a well established method in biomedical research \cite{1, 2}. In addition to confocal laser scanning microscopy (CLSM), where the spatial fluorescence intensity distribution is sampled, FCS records the time-dependent intensity fluctuations originating from a microscopic spot. With sufficient sensitivity and time resolution of the detectors the signal fluctuations will reflect not only instrumental noise but also the stochastic Brownian motion of the fluorescent molecules performing random walks through the illuminated, open detection volume. Thus, the time-dependent signal contains information about molecular properties which can be extracted and analysed \cite{weide}.

The average time required by molecules to ‘walk’ through the detection volume (diffusion time) depends on its shape and size and is therefore characteristic for the optical setup. The mobility of the molecules depends on their shape and size, as well as the viscosity of the surrounding medium \cite{weide}. Because the viscosity of a lipid bilayer or a cellular plasma membrane is about 100 times larger than that of the aqueous media, the diffusion times are slowed down to the same degree. In addition, the diffusion is confined to a very thin membrane plane of 5-10 nm thickness. Using diffraction limited illumination, lateral, two-dimensional diffusion times of lipids are typically in the millisecond (ms) regime and single-pass transmembrane receptors are even 20-50 times slower.

\begin{figure}
\begin{center}
\includegraphics[width=.8\linewidth]{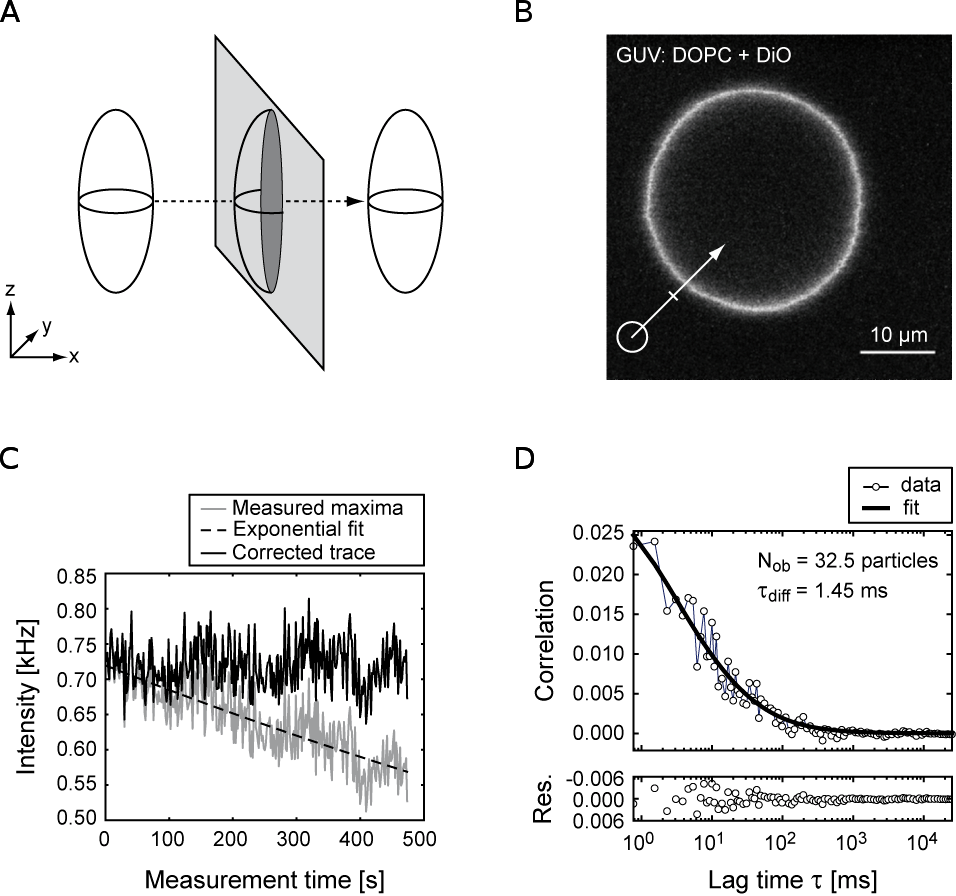}
\end{center}
\caption{Single-focus perpendicular SFCS}{
(\textbf{A}) Graphical representation of the open detection volume as an ellipsoid, stretched along the optical axis z. Lateral movement on a perpendicular scan path (dashed arrow) through a two-dimensional membrane plane (light grey) creates an elliptical, two-dimensional detection area (dark grey). (\textbf{B}) Equatorial confocal image of a GUV composed of DOPC containing 0.007\,\% of the fluorescent lipid DiO. The scan path (white arrow) was positioned perpendicular to the membrane plane at a 45° angle. (\textbf{C}) Intensity trace of the maxima (grey) along each scan is plotted against the measurement time (500 s = 6.5$\times$105 scans with a cycle time \SI{0.768}{ms}). During the measurement, the intensity dropped exponentially (dashed) by about 25\,\% due to bleaching, and was back-corrected (black). The magnitude of the intensity values (here about \SI{0.75}{kHz}) depends on the exact procedure to assign membrane related fluorescence, in particular the number of bins used to sample each line scan (here 70) and the number of bins assigned to the membrane area (here seven: 1 maximum plus six adjacent bins). (\textbf{D}) Autocorrelation curve of the corrected intensity trace from C was fitted with a model function for two-dimensional diffusion through an elliptical Gaussian shaped detection volume (equation~\ref{eq2}).\label{fig1}}
\end{figure}

Perpendicular scanning (SFCS) was introduced to measure mobility and interactions in the lipid bilayer of giant unilamellar vesicles (GUVs) \cite{3,4}. Slower diffusion of fluorescent molecules in combination with small concentrations produce rare events with longer lifetime and therefore it is sufficient to collect data with an appropriate sampling frequency. In perpendicular line SFCS \cite{3, 5, 6}, this is achieved by guiding the focused laser beam along a straight line with kHz frequency perpendicular through the membrane plane (figure~\ref{fig1}). Scanning is performed by means of the conventional control software of a CLSM. At all times, the emitted photons are collected with single photon counting detectors, e.g. avalanche photo diodes (APDs). The signal is digitalized by a separate device and stored in a file. Here we used a USB connected hardware correlator, however other time-correlated single photon counting (TCSPC) cards may also be suitable. Since the file contains all photon incidents along the scan path, the signal originating from the spot where the scan path crosses the membrane plane must be extracted post-measurement.

For this purpose, we developed a program (\textit{PyScanFCS}). The correct periodicity (scan cycle time) corresponding to the repetition rate of each scan is determined by means of Fourrier analysis of the photon arrival times. For visual inspection, the aligned scan intensities are displayed in a kymograph containing the number of scans on the x-axis and the intensity along the scan path on the y-axis. The time range displaying homogeneous intensity can be manually selected. The number of bins along the scan path containing membrane related intensity can be defined
 and corrected for systematic bleaching during the course of the measurement (typically several minutes). Finally, the resulting intensity trace is time-correlated by a multiple-\textgreek{τ} algorithm. To evaluate the correlation data, the curves can be analysed with a second program (\textit{PyCorrFit}) written for fitting different model functions containing molecular parameters like diffusion times, diffusion coefficients, and particle numbers \cite{Muller_2014}.

Compared to single-point confocal FCS, perpendicular SFCS has several advantages when applied to membranes \cite{5, 7}. Free standing membranes are usually large structures showing slow thermal motion like undulations. During the measurement, the vesicle or the cell may move or change shape. This is a problem for conventional single-point confocal FCS, because the detection volume is fixed in space; even minor displacements of the fluorescent membrane with respect to a fixed detection volume cause a detrimental bias in correlation curves. Because in SFCS the maximum intensity on each path is determined post-measurement, such positional instabilities can be corrected. Additionally, the laser focus will be outside the membrane most of the time and therefore the amount of bleaching is reduced. However, a potential disadvantage of the SFCS approach is a low average signal-to-noise-ratio and accordingly prolonged measurement times.

We demonstrate how to obtain and evaluate perpendicular SFCS data by measuring the diffusion of the 488 nm excitable, fluorescent lipid 3,3'-Dioctadecyloxacarbocyanine (DiO) in GUVs composed of 1,2-dioleoyl-sn-glycero-3-phosphocholine (DOPC). Scanning is either performed on a single line (single-focus SFCS) or on a frame composed of two lines (two-focus SFCS). Two-focus SFCS, takes advantage of the fixed distance between the scan paths to obtain calibrated diffusion coefficients \cite{8}. Additional scan modalities like dual-colour scanning fluorescence cross-correlation spectroscopy (SFCCS) \cite{9, 10} are briefly outlined. Dual-colour SFCCS is a powerful approach to address interactions between differently labelled chemical species in biomembranes.

\section{Materials}
\subsection{Preparation of GUVs}
\begin{enumerate}
\item \SI{10}{ml} \SI{200}{mM} sucrose in H2O
\item \SI{10}{ml} \SI{200}{mM} glucose in H2O
\item Preparation of lipid mixture: Take DOPC stock solution (\SI{20}{mg/ml} in chloroform) and DiO (\SI{5}{\micro M} in chloroform) from -20°C, and dilute with chloroform to final concentration of \SI{2}{mg/ml} DOPC and \SI{0.007}{mol \%} DiO (Avanti Polar Lipids, Alabaster, Alabama, USA). 5-\SI{10}{\ul} are used to prepare GUVs in the Teflon chamber.
\item Bovine serum albumin (BSA), \SI{2}{mg/ml} in H2O (Sigma-Aldrich, Munich, Germany)
\item Vacuum chamber for drying samples (Vacuubrand, Wertheim, Germany).
\item Home made Teflon chamber containing two platinum electrodes; capacity \SI{300}{\ul} (see note~\ref{6.1})
\item Electrical sine wave generator (TTi TG315 function generator, Huntingdon, Cambridgeshire, UK)
\end{enumerate}

\subsection{Microscopy}
\begin{enumerate}
\item 1. 8-well chambered cover slides (LabTek II, $\#$ 1.5, Nunc, Thermo Scientific, USA).
\item Diffusion standard to calibrate the detection volume: hydrolized AlexaFluor 488 (Life technologies/Molecular Probes, USA). Prepare a \SI{25}{nM} solution in \SI{10}{mM} Tris/H2O, pH 8.
\item Purified enhanced green fluorescent protein (eGFP) (CatNo.: $\#$4999-100, BioVision, USA).
\end{enumerate}

\subsection{Hardware for SFCS data acquisition}
CLSM setups equipped with photon counting detectors (e.g. a LSM510 with accessory Confocor3 unit, Zeiss, Germany) offer direct access to the signal by means of BNC connected coaxial cables attached to the APDs. If the CLSM does not possess photon counting detectors, one can use the fibre-out exit and feed the signal into external APDs \cite{3}. For this, one needs
\begin{enumerate}
\item  APDs (Perkin-Elmer Optoelectronics, USA)
\item  Multimode fibres with MM connectors, ASF105/125Y (Thorlabs, Newton, New Jersey, USA)
\item  Positioning system (miniature xyz stage) with FC adapter to align the optical fibre (Owis, Staufen, Germany)
\item  USB connectable hardware correlator to record the photon history (Model Flex02-12D/B, \url{www.correlator.com}, USA/China) (see note~\ref{6.2}).
\end{enumerate}

\subsection{Software for SFCS data acquisition}
The data processing software (\textit{PyScanFCS} and \textit{PyCorrFit}) has been tested on Windows, Debian/Ubuntu Linux, and Mac OSx (see note~\ref{6.3}).
\begin{enumerate}
\item Software for photon history recording, \texttt{Photon.exe} (shipped with the Flex correlator Flex02-12D/B, \url{www.correlator.com})
\item Program to generate correlation curves from the photon history file \textit{PyScanFCS} (free download at \url{http://pyscanfcs.craban.de}). Due to constraints in memory of 32-bit systems, we recommend to use a 64-bit computer with either Windows 7 or an Ubuntu-amd64 operating system.
\item Fitting program to extract relevant parameters from correlation curves, \textit{PyCorrFit} (free download at \url{http://pycorrfit.craban.de}).
\end{enumerate}

\section{Methods}
\label{5}
\subsection{Preparation of GUVs by electroformation}
\label{5.1}
\begin{enumerate}
\item Clean the cap and reservoir of the Teflon chamber two times with \SI{80}{\%} Ethanol and with hot tap water.
\item Rinse the Teflon container and the cap with distilled water and dry both with pressurized air.
\item Using a pipette, distribute a total \SI{5}{\ul} of the \SI{2}{mg/ml} lipid-dye mixture evenly on both platinum electrodes pointing towards the inside of the chamber.
\item To remove the chloroform, let the lipid mixture dry in air for \SI{2}{min} and further in a vacuum chamber for \SI{10}{min}.
\item Fill the Teflon container with \SI{300}{\ul} of \SI{200}{mM} sucrose solution and close the lid. The platinum electrodes should be completely covered by sucrose solution.
\item Apply an alternating electric field of \SI{2.15}{V} at \SI{10}{Hz} for \SI{1.5}{h}. During this phase the GUVs form at the electrodes.
\item Decrease the frequency to \SI{2}{Hz} for another \SI{20}{min} to allow the GUVs to detach from the electrodes.
\item Incubate the Labtek chamber with \SI{2}{mg/ml} bovine serum albumin (BSA) for \SI{20}{min}.
\item Wash the Labtek chamber five times with \SI{200}{mM} glucose solution.
\item Fill the Labtek chamber with \SI{400}{\ul} of \SI{200}{mM} glucose solution.
\item Add \SI{50}{\ul} of the GUV-sucrose solution. Cut off the tip when pipetting GUVs to prevent them from bursting due to shear forces (see note~\ref{6.4}).
\end{enumerate}

\subsection{Define the microscope settings for SFCS}
\label{5.2}
\subsubsection{Calibration with Alexa488 solution}
\label{5.2.1}
\begin{enumerate}
\item Prepare the 8-well Labtek containing a 1-\SI{50}{nM} Alexa488 solution and the GUV solution in adjacent wells. Clean the glass bottom with ultra pure EtOH and mount on the stage.
\item Define settings for excitation and detection (e.g. a dichroic 488/633 followed by BG35 to eliminate scattered UV light and a cleaning band pass (505-560) for Alexa488 and DiO emission).
\item Set the CLSM to measure in a single spot (with LSM510 and external APDs this is possible by selecting scan mode ‘Spot’ and performing a time series).
\item Position the focus in solution at \SI{50}{\um} above the glass bottom (see note~\ref{6.5}).
\item Adjust laser power to 50-\SI{100}{kHz} and maximize the count rate at the APD by iterative alignment of pinhole position (by means of the LSM software, Confocor3) and the external fibre coupling (by moving the external miniature xyz stage).
\item Adjust the correction collar of the objective to the thickness of the Labtek cover glass by maximizing the count rate.
\item Record three autocorrelation curves (1-\SI{2}{min}) at three different lateral positions and save as a \mytilde .sin file (Flex correlator software) or \mytilde .fcs file (Confocor3).
\item Evaluate the data with \textit{PyCorrFit} using a confocal 3D-diffusion model with triplet contribution to obtain the average diffusion time and the axis ratio (‘structure parameter’) of the detection volume. The residuals of the fit should not show a systematic bias indicating proper alignment and nearly Gaussian geometry of the point spread function. Calculate the effective focal volume (see note~\ref{6.6}).
\end{enumerate}

\subsubsection{Calibration for two-focus SFCS}
\label{5.2.2}
\begin{figure}
\begin{center}
\includegraphics[width=.9\linewidth]{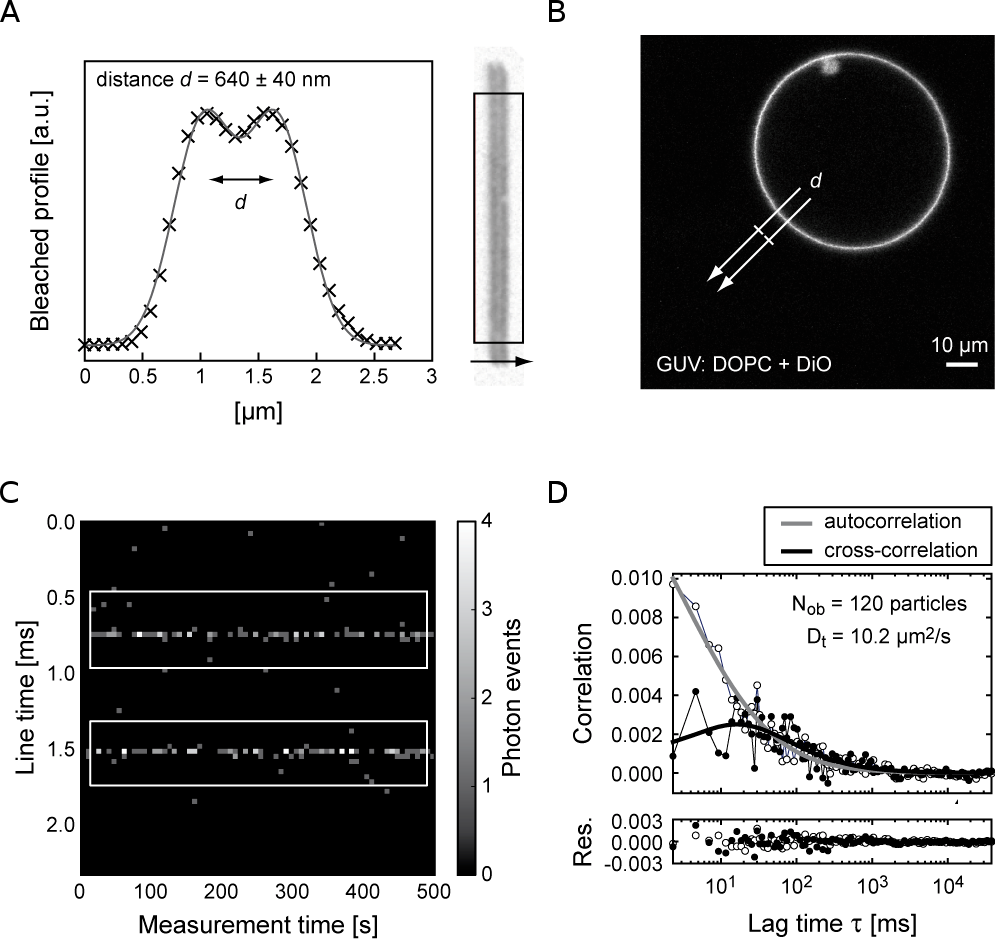}
\end{center}
\caption{Two-focus perpendicular SFCS}{(\textbf{A}) Calibration procedure. The image shows eGFP protein adsorbed to the glass surface after continuous scanning for about 1 min (a reference pre-bleach image was subtracted). The two dark lines representing the parallel scan path were integrated (rectangular area, ImageJ) and the profile fitted with equation~\ref{eq1} to derive the mean distance $d$. (\textbf{B}) Equatorial confocal image of a GUV composed of DOPC containing 0.007\,\% of the fluorescent lipid DiO. The scan path (white arrow) was positioned perpendicular to the membrane plane at a 45° angle. (\textbf{C}) Kymograph containing the binned (70) photon counts along the scan path (y-axis, in total \SI{2.3}{ms}) during the measurement time (x-axis, in total \SI{500}{s}). Since the focal volume passes the membrane twice during each scan cycle, two horizontal lines appear. (\textbf{D}) Averaged autocorrelation and cross-correlation curves were fitted with equation~\ref{eq2} and \ref{eq3}, respectively. The diffusion coefficient was calculated by applying equation~\ref{eq4}.\label{fig2}}
\end{figure}

To take advantage of the calibration-free determination of diffusion coefficients with two-focus SFCS, the signal from two parallel scan paths is cross-correlated \cite{8} (figure~\ref{fig2}). However, to evaluate the cross-correlation curves, one needs to know their physical distance d. This can be determined by bleaching a layer of eGFP adsorbed to the glass bottom of the well and fitting a double Gaussian to the intensity profile (figure~\ref{fig2}A):
\begin{equation}
P(d) = \frac{A}{\sqrt{2 \pi}\sigma} \left[ 
			e^{-\frac{1}{2}\left( \frac{x-\mu}{\sigma} \right)^2} 
			e^{-\frac{1}{2}\left( \frac{x-\mu - d}{\sigma} \right)^2} 
			\right] + B
\label{eq1}
\end{equation}
where $x$ is a spatial coordinate normal to the scan path, $\mu$ is the location of the first maximum, $\sigma$ the standard deviation, and $d$ is the distance between the maxima. The parameters $A$ and $B$ describe the amplitude and offset of the cross-section. Because of hysteresis effects, not every scanning procedure reproduces two parallel beam paths (see note~\ref{6.7}). A sufficiently accurate two-focus scan could be achieved by the following sequence (AIM software, Zeiss):
\begin{enumerate}
\item Scan mode ‘Multitrack’ with the 1. track laser off and the 2. track laser on; change tracks between each scanned line.
\item Bi-directional scanning.
\item Zoom 12, angle: 45°.
\item Acquisition: Frame mode 32x2 pixels.
\end{enumerate}
By this, the first line is scanned forward without laser. The second line is scanned backwards with the laser on and fluorescence is detected (first detected scan). The second line is again scanned, but forwards, without laser. The scan cycle is completed with a scan of the first line backwards while the laser is on (second detected scan). As seen in the bleached image (figure~\ref{fig2}A, right panel), there is only minimal hysteresis at the beginning and at the end of the scan path.

\subsection{Data acquisition: One-focus SFCS}
\label{5.3}
\begin{enumerate}
\item Perform Alexa488 calibration according to section~\ref{5.2.1}
\item Use the CLSM software to image the equatorial plane of a GUV.
\item Switch to line scanning and ‘crop’ mode, and define a scan path perpendicular to the membrane (in the LSM) (see note~\ref{6.8}).
\item Set the scan mode to ‘unidirectional’ at maximum scan rate (e.g. LSM510, \SI{0.786}{ms} at ‘zoom 12’).
\item Direct the emission light to the APD (Confocor) or via the fibre-out exit to the external detector.
\item Execute the software \texttt{Photon.exe} to record the photon stream. Choose the file format \SI{16}{bit}. Set the acquisition time between \SI{200}{s} and \SI{500}{s}. Set the channels to “A only” since this protocol uses only one APD. Make sure that the APD is connected to the “input A” of the Flex correlator.
\item To measure, first start a continuous line scan in the LSM software, then start file recording in \texttt{Photon.exe}.
\end{enumerate}
The program \texttt{Photon.exe} creates a \mytilde .dat file (see note~\ref{6.9}) that contains the entire photon history of the measurement. The file has to be processed using \textit{PyScanFCS}, described in section~\ref{5.6}.

\subsection{Data acquisition: Two-focus SFCS}
\label{5.4}
\begin{enumerate}
\item Perform Alexa488 calibration according to section~\ref{5.2.1}
\item Use the CLSM software to image the equatorial plane of a GUV.
\item Apply the exact measurement settings (zoom, scanning speed, scanning angle) that were recorded during the calibration of the distance d between the scanned lines in section~\ref{5.2.2}. Position the scan path in the image (see note~\ref{6.8}).
\item Direct the emission light to the APD (Confocor) or via the fibre-out exit to the external detector.
\item Execute the software \texttt{Photon.exe} to record the photon stream. Choose the file format \SI{16}{bit}. Set the acquisition time between \SI{200}{s} and \SI{500}{s}. Set the channels to “A only” since this protocol uses only one APD. Make sure that the APD is connected to the “input A” of the Flex correlator.
\item To measure, first start continuous frame scanning in the LSM software, then start file recording in \texttt{Photon.exe}.
\end{enumerate}
The program \texttt{Photon.exe} creates a \mytilde .dat file (see note~\ref{6.9}) that contains the entire photon history of the measurement. The file has to be processed using \textit{PyScanFCS}, described in section~\ref{5.6}.

\subsection{Data acquisition: Dual colour applications}
\label{5.5}
In dual-colour SFCCS, the emission pathway is separated by an appropriate beam splitter and the signal is detected by two APDs. The scanning settings to measure dual-colour SFCS are the same as for ‘One-focus SFCS’ as described in section~\ref{5.3}, with the extension that both laser lines and their corresponding detection pathways are simultaneously used. For ‘alternating excitation’ the colour channels are scanned sequentially (LSM510, ‘Multitrack mode’). Since two APDs are now connected to the Flex correlator in the program \texttt{Photon.exe}, channels A and B have to be selected for data recording. SFCCS with alternating excitation is a convenient way to circumvent crosstalk between the colour channels, for example when observing green and red fluorescent proteins \cite{9, 10}.

\subsection{Data processing with PyScanFCS}
\label{5.6}
\begin{figure}
\begin{center}
\includegraphics[width=.7\linewidth]{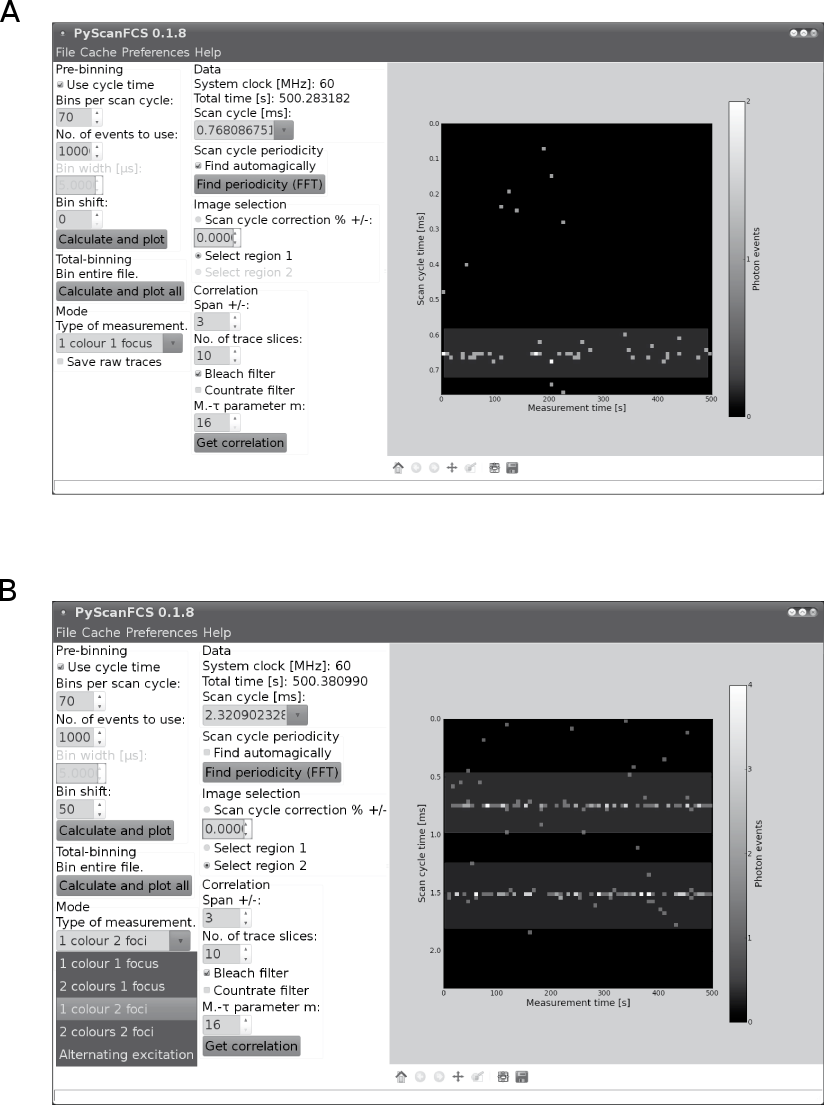}
\end{center}
\caption{\textit{PyScanFCS} user interface}{
\textit{PyScanFCS} is a tool to generate correlation curves from a stored photon stream. (\textbf{A}) Screenshot showing the ‘1-colour-1-focus’ application when evaluating a single-focus FCS measurement. The complete file (\SI{500}{s}) was loaded and binned (70). By means of FFT, the program determined a cycle time of \SI{0.768}{ms}. Prior to calculating the correlation functions, the horizontal intensity trace was selected by dragging a rectangle (grey shaded). (\textbf{B}) Example for ‘2-colour-1-focus’ application (see pull-down menu for other measurement modes). ‘Bin shift’ (50) was used to place the horizontal intensity traces in the centre of the kymograph window.
\label{fig3}}
\end{figure}

\textit{PyScanFCS} can extract membrane related intensity fluctuations from the photon history of the scan and calculate experimental correlation functions (figure~\ref{fig3}).
\begin{enumerate}
\item Loading the data. Open an experimental data file (File $|$ Open photon events .dat file). In the ‘Pre-binning’ box, set the bins per scan cycle to 70 and the number of used photon events to 1000 (see note~\ref{6.10}). Set the bin width to \SI{5}{\us} and click ‘Calculate and plot’.
\item The scan cycle time needs to be found. In the ‘Scan cycle periodicity’ box check ‘Find automatically’ and click ‘Find periodicity (FFT)’. The program performs a Fourier transform of the photon arrival times on the first 1000 events and searches for a maximum amplitude. Check the value of the cycle time found by \textit{PyScanFCS}. If this value does not match the magnitude as expected from the microscope settings, uncheck ‘Find automatically’ and click ‘Find periodicity (FFT)’ again. A new interactive window pops up displaying all amplitudes of the FFT and their corresponding cycle times. Manually select the expected cycle time by clicking on both sides of the peak (note~\ref{6.11}).
\item Binning of the entire file. In the ‘Pre-binning’ box, check the ‘Use cycle time’ option. The displayed cycle time will be used to calculate the kymograph. The plot on the right hand side should show a line of bins which represents the intensity trace measured in the membrane. If this line is tilted, the cycle time has to be corrected: Select ‘Scan cycle correction’ in the ‘Image selection’ box and manually drag a line along the intensity trace inside the kymograph window. Clicking again ‘Calculate and plot’ will reproduce a corrected, horizontal line. Increase the ‘No. of events to use’ step-wise by a factor of 10 and subsequently correct the cycle time until the entire \mytilde .dat file has been processed, i.e. the total measurement time has been plotted.
\item Trace selection in the kymograph. If one-focus SFCS was performed, a single trace should be visible in the kymograph. If two-focus SFCS was performed, two traces representing the two parallel scan paths should be visible. In the “Mode” box, select ‘1-colour-1-focus’ or ‘1-colour-2-foci’ according to your measurement. In the ‘Image selection’ box, choose ‘Select region 1’ and draw a rectangle comprising a region of homogeneous intensity. For two-focus SFCS ‘Select region 2’ is used to manually select the second trace.
\item Calculation of the intensity trace and the correlation curve. In the ‘Correlation’ box, define the number of neighbouring bins along the scan path that will be added up to the final number of counts for this time point. With a resolution of 70 bins per scan cycle, a ‘span’ of ±3 is a good measure. The trace can be split in a number of slices each of which will be correlated separately. This is useful when inhomogeneities like temporary signal loss or bright aggregates appear during the measurement. In \textit{PyCorrFit}, the correlation curves containing the corresponding bias can be sorted out (see note~\ref{6.12}). Check ‘Bleach filter’ if you want to correct for bleaching during the measurement \cite{6}. \textit{PyScanFCS} is able to perform an approximation of the actual count rate using the ‘Countrate filter’ (not done here). The real count rate at the membrane is about two orders of magnitudes greater than shown in figure~\ref{fig1}C. The magnitude of the intensity does not affect the calculation of the correlation curves.  The theoretical background of the count rate filter is described in the \textit{PyScanFCS} documentation. The parameter m of the multiple-\textgreek{τ} algorithm should be set to 16. Finally, click on ‘Get correlation’. This will calculate the correlation curves and save them as a \mytilde .zip file inside the measurement folder.
\end{enumerate}

\subsection{Data processing with PyCorrFit}
\label{5.7}
\begin{figure}
\begin{center}
\includegraphics[width=.5\linewidth]{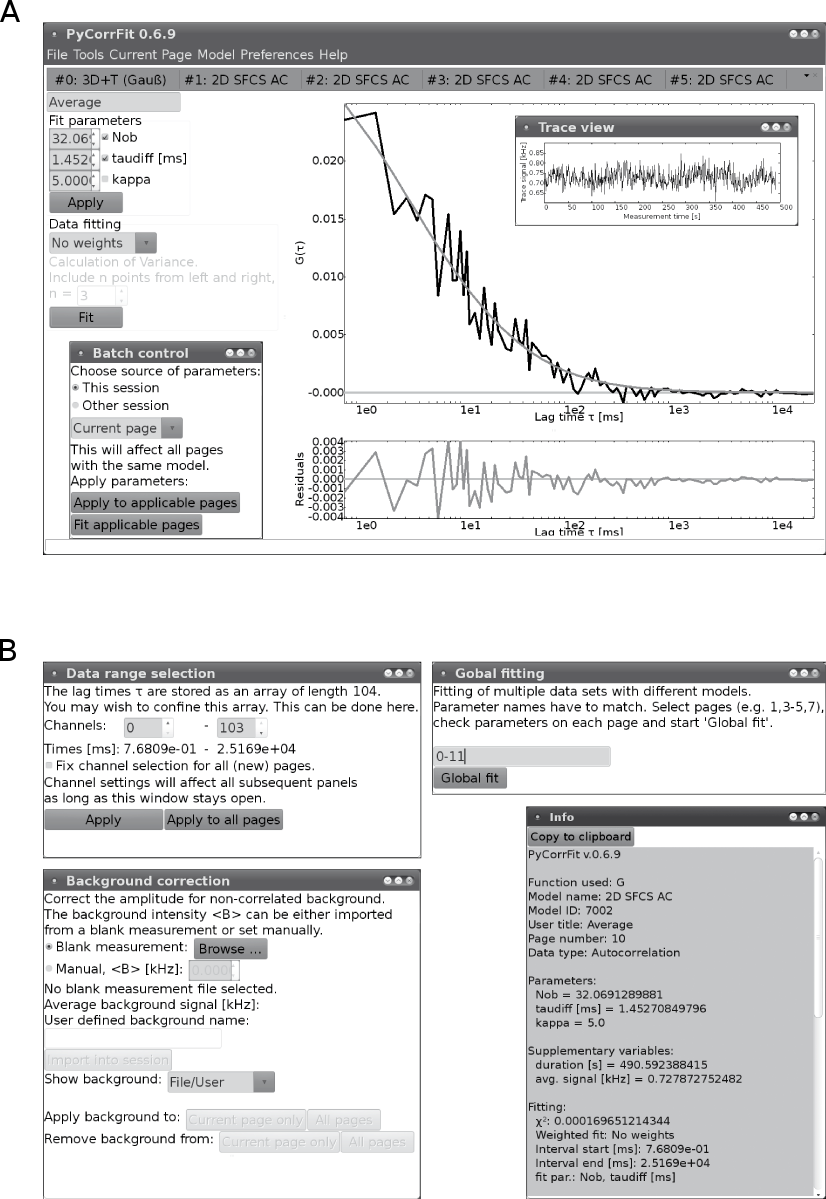}
\end{center}
\caption{\textit{PyCorrFit} user interface}{
\textit{PyCorrFit} is a tool to evaluate correlation functions. (\textbf{A}) Main window showing the menu bar, tabs containing experimental or averaged correlation curves and the fitting parameters. The set of fitting parameters is defined by the model function which must be loaded beforehand. The experimental correlation data (black) and the actual fit function (grey) are shown in a semi-log plot together with the residuals displaying the difference between the data and the fit. Note that ‘Trace view’ displays the corresponding intensities and allows for removing outliers. With ‘Batch control’ the user can act on multiple panels and therefore allow controlled fitting of large data sets. (\textbf{B}) Examples for additional functions: ‘Data range selection’ restricts the number of time channels of the correlation data as used for fitting, ‘Background correction’ accounts for non-correlated intensity to determine absolute concentrations, and ‘Global fit’ allows for simultaneous fitting of data sets containing shared parameters in their model functions. At any time, the user can call information panels specifying the current fitting parameters and models.
\label{fig4}}
\end{figure}

\textit{PyCorrFit} can fit analytical model functions to experimental correlation curves (figure~\ref{fig4}). For measurements in GUVs, a two-dimensional diffusion model was applied
\begin{align}
G(\tau) = \frac{1}{N_\text{ob}} 
\left( 1+\frac{\tau}{\tau_\mathrm{diff,2D}} \right)^{-1/2}
\left( 1+ \frac{\tau}{\kappa^2 \, \tau_\mathrm{diff,2D}} \right)^{-1/2}
\label{eq2}
\end{align} 
containing the diffusion time $\tau_\mathrm{diff,2D}$, the particle number $N_\text{ob}$, and the axis ratio $\kappa$ of the ellipsoid shaped detection area as parameters. For two-focus SFCS, a cross-correlation function can be calculated and fitted to determine the distance between the parallel scan paths $d$ and the radius of the beam waist $w_0$ 
\begin{align}
G\!_\times(\tau) = G(\tau) \cdot \exp\!\left[ \frac{-d^2}{w_0^2 (1+\tau/\tau_\text{diff,2D})} \right] 
\label{eq3}
\end{align}
The beam waist $w_0$ is needed to convert diffusion times into diffusion coefficients
\begin{align}
D^{\text{2D}} = \frac{w_0^2}{4\tau_\text{diff,2D}}.
\label{eq4}
\end{align}
Evaluation of the correlation data comprises the following steps:
\begin{enumerate}
\item Import the appropriate model function via ‘File $|$ Import model function’ (see note~\ref{6.13}). Import
 \texttt{ExampleFunc{\_\-}SFCS{\_\-}1C{\_\-}2D{\_\-}Autocorrelation.txt} for one-focus SFCS (Equation 2) and  import \texttt{ExampleFunc{\_\-}SFCS{\_\-}1C{\_\-}2D{\_\-}Cross-correlation.txt} for two-focus SFCS (Equation 3). The imported autocorrelation and cross-correlation model functions are now available in the ‘Model $|$ User’ menu, depicted as ‘2D SFCS AC’ and ‘2D SFCS CC’, respectively. 
\item Import the \mytilde .zip file containing the correlation data created by \textit{PyScanFCS} via ‘File $|$ Load single data file’. As a model select the previously imported model function ‘User: 2D SFCS AC’. For two-focus SFCS, you will be asked which correlation curves to import. First choose the autocorrelation (AC) curves and select the model ‘User: 2D SFCS AC’. The cross-correlation (CC) curves need to be imported separately with the model ‘User: 2D SFCS CC’. Note that the data import function associates each correlation curve with a particular model.
\item Now several tabs are shown, each containing a correlation curve corresponding to the number of intensity traces. Open ’Tools $|$ Data range selection’ to eventually narrow the window of lag times as used for fitting. Eventually apply a ‘Background correction’ for non-correlated intensity. Check parameters (e.g. $\tau_\text{diff}$ and $N_\text{ob}$) that should be varied during the fitting procedure and start by clicking ‘Fit’. Use ‘Tools $|$ Batch control’ to apply the same settings when fitting multiple curves (tabs) assigned to the same model. Browse through the results while checking the intensity traces with ‘Tools $|$ Trace view’ in a separate window. Correlation curves with unstable intensities can be removed by closing the tab.
\item Using ‘Tools $|$ Average data’, multiple correlation curves can be averaged; the new curve will be shown in a new tab and can be fitted separately. To evaluate two-focus SFCS data, it is necessary to create an average for both AC and CC functions and save each in a separate \mytilde .csv file by ‘Current Page $|$ Save data (*.csv)’. Clear the session with ‘File $|$ Clear session’ and re-import the saved \mytilde .csv files with their corresponding model functions ‘2D SFCS AC’ and ‘2D SFCS CC’ via ‘File $|$ Load single data file’. Since autocorrelation and cross-correlation curves share parameters, a global fit can be applied. Activate the ‘Tools $|$ Global fitting’ menu. In each tab, check the parameters that should be varied during the fitting procedure ($\tau_\text{diff}$, $N_\text{ob}$, $w_0$) and start by clicking on ‘Gobal fit’. The fit results are then used to calculate the diffusion coefficient according to Equation (4).
\item Fit result for each panel as well as for the average curves are displayed in the main window. In addition, you can use ‘Tools $|$ Page info’, which contains a comprehensive parameter set for each fit. Finally, you can save the entire session by ‘File $|$ Save session’ or each individual correlation curve via ‘Current Page $|$ Save data (*.csv)’. In both cases, the fit parameters will be saved as well. More information and a full documentation of \textit{PyCorrFit} are available online at \url{http://pycorrfit.craban.de}.
\end{enumerate}

\section{Notes}
\label{6}
\subsection{GUV teflon chamber}
\label{6.1}
The cylindrical chamber (\SI{1}{cm} diameter, \SI{2}{cm} height) is made out of a Teflon cap which can be screwed onto a reservoir. The cap is pierced by a pair of platinum electrodes protruding \SI{1}{cm} on each side. The distance between the wires is \SI{4}{mm}. The platinum wires can be connected to a power supply and are used to generate an electric field inside the chamber.

\subsection{Correlator model types}
\label{6.2}
Unfortunately, the photon history recorder mode does not work with the newer correlator version Flex02-01D/C. Please consult \url{www.correlator.com} or use other time correlated single photon counting (TCSPC) cards and convert the data accordingly (see note~\ref{6.9}).

\subsection{Obtaining PyScanFCS and PyCorrFit}
\label{6.3}
The two software packages, \textit{PyCorrFit} and \textit{PyScanFCS}, enable the user to derive molecular parameters from an SFCS experiment. The program \textit{PyScanFCS} is used to extract experimental correlation curves from the photon stream recorded during scanning. The correlation curves are stored on the hard disk and can be evaluated by \textit{PyCorrFit}, a generic front-end tool for fitting of model functions. \textit{PyCorrFit} contains additional features exceeding SFCS applications that are described in detail in the documentation. Both programs are available from our download page (\url{http://fcstools.craban.de}). The source code (Python) and the executables are free software published under the GNU general public licence. There are also pre-built binaries/executables for Windows, Mac OSx, and Ubuntu-Linux.

\subsection{Osmolarity and density of GUVs}
\label{6.4}
When the osmolarity outside and inside of the GUVs differ, the GUVs are instable and may burst. Therefore the molaritiy on both sides of the vesicle membrane should be the same. Due to the larger molecular weight of sucrose (inside) than glucose as contained in the glucose/sucrose mixture (outside), the negative buoyancy of the GUVs enforces their sedimentation to the glass bottom where they can be conveniently observed with an inverted microscope.

\subsection{Determination of the axial measurement position}
\label{6.5}
Turn the course and fine focus to drive the detection volume from the glass bottom into solution and verify the sudden increase in signal intensity. At the glass bottom, reset the position offset to zero, then move \SI{50}{\um} up.

\subsection{Computation of diffusion coefficients}
\label{6.6}
To transform diffusion times into diffusion coefficients one has to calculate $D=D_\text{st} \cdot \tau_\text{diff,st} /\tau_\text{diff} $ from the measured diffusion time $\tau_\text{diff}$ and the standard $\tau_\text{diff,st}$ for which the diffusion coefficient $D_\text{diff}$ is known. The diffusion coefficient for Alexa488 in \SI{10}{mM} Tris pH 8 measured at 22°C is (4.35 ± 0.1)$\times 10^{-10}\,$\SI{}{m^2\,s^{-1}}
 \cite{11}. The focal volume can be derived by applying $V_\text{eff} = \pi^{3/2} \kappa w_0^3$. Note that the axis ratio and the diffusion time trade during the fit and thus the axis ratio is usually a poorly determined parameter. To achieve consistency, it is quite common to determine $\kappa$ in solution for the standard like Alexa488 ($3 < \kappa < 8$) and to fix this value for all measurements.

\subsection{Scan path alignment}
\label{6.7}
The scan path should be checked for several angles. For example, the parallel arrangement was not maintained when scanning horizontally at 0°. Furthermore, our chosen settings in ‘multitrack mode’ are obviously not optimal in terms of time resolution. Using other instruments, one may achieve hysteresis-free bidirectional scanning with correspondingly shorter scan cycle times.

\subsection{Scan path through a GUV}
\label{6.8}
The centre of the scanning line should be positioned close to the membrane but outside of the GUV. Since the GUV membrane bends inwards, excitation and bleaching above and below the equatorial plane is therefore minimized.

\subsection{PyScanFCS data file formats}
\label{6.9}
\textit{PyScanFCS} can only import \mytilde .dat files or pre-binned \mytilde .fits files (see online documentation at \url{pyscanfcs.craban.de}). \mytilde .dat files are created by the software \texttt{Photon.exe} that ships with the correlators Flex02-12D/B. If you are using a different photon history recording device, you will have to transform your data files to the \mytilde .dat file format. The format stores the time span between two photon events in units of the system clock. The data format is as follows: 
\begin{itemize}
\item[-] The value of the first byte (\texttt{uint8}) identifies the format of the file: 8, 16, or \SI{32}{bit}. 
\item[-] The second byte (\texttt{uint8}) identifies the system clock in MHz, usually 60. 
\item[-] The rest of the file contains the time span between two photon events. The time unit is 1/\SI{60}{MHz} = \SI{16.67}{ns}. 
\item[-] \SI{8}{bit} format is not supported.
\item[-] \SI{16}{bit} format. Each \texttt{word} (2 bytes, \texttt{uint16}) represents a photon event. For low intensities, values of \texttt{0xFFFF} may occur. In this case, the following four bytes (\texttt{uint32}) represent a photon event. 
\item[-] \SI{32}{bit} format. Each \texttt{dword} (4 bytes, \texttt{uint32}) represents a photon event. Can be created from \SI{16}{bit} files using \texttt{Photon.exe}. \textit{PyScanFCS} saves \mytilde .dat files in this format.\end{itemize}

\subsection{Binning photon stream data}
\label{6.10}
The binning resolution is an empirical value balancing computation time and spatial scanning resolution. Under conditions as used here, membrane related photon events usually appear in 1-3 adjacent bins. Initially limiting the number of photon events to 1000 speeds up the determination of the cycle time by the fast Fourier transform (FFT).

\subsection{Determination of scan cycle time}
\label{6.11}
In the FFT spectrum the maximum amplitude does not always reflect the correct scan cycle time. Moreover the scan periodicity as displayed by the LSM software may differ from the real physical value.

\subsection{Scan time resolution}
\label{6.12}
The remaining time span of the split traces should still exceed the diffusion times by at least a factor of 1000.

\subsection{Fitting user-defined models}
\label{6.13}
\textit{PyCorrFit} can import external, user defined model functions. An external model function must be defined and stored in a \mytilde .txt file with a simple syntax. Examples and instructions on how to create such a model functions can be found online (\url{http://pycorrfit.craban.de}). As in standard confocal SFCS, here we utilize well-known autocorrelation functions derived for Gaussian detection profiles.

\section*{Acknowledgments}
We thank Fabian Heinemann and Eugene Petrov for helpful discussions.

\bibliographystyle{ieeetr}

\bibliography{pplsFCS} 

\end{document}